\documentclass{ph2}
\usepackage{txfonts,natbib}
\usepackage{graphicx}

\begin{document}
\begin{frontmatter}
\title{Attractiveness and activity in Internet communities}
\author{Gourab Ghoshal, Petter Holme\thanksref{auth:ph}}
\thanks[auth:ph]{Corresponding author. \textit{Tel.:} +1 734 647 9568,
  \textit{E-mail:} pholme@umich.edu}

\address{Department of Physics, University of Michigan, Ann
  Arbor, MI 48109, U.S.A.}
\begin{abstract}
  Datasets of online communication often take the form of contact
  sequences---ordered lists contacts (where a contact is defined as a
  triple of a sender, a recipient and a time). We propose measures of
  attractiveness and activity for such data sets and analyze these
  quantities for anonymized contact sequences from an Internet dating
  community. For this data set the attractiveness and activity
  measures show broad power-law like distributions. Our attractiveness
  and activity measures are more strongly correlated in the real-world
  data than in our reference model. Effects that indirectly can make
  active users more attractive are discussed.
\end{abstract}

\end{frontmatter}

\section{Introduction}

The Internet has revolutionized science and society in many
ways. Studies of human communication networks nowadays deal with data
sets of thousands people or more. Gone are the days when one had to rely
solely on interview surveys and observational studies. These new and
larger data sets open up new possibilities---one does not only get higher
precision in the measurement of network structure, one can also rid
the structural quantities of finite size effects that would hinder the
prediction of phenomena such as large-scale information spreading. The
accessibility of large data sets has also drawn the interest of
statistical physicists, habitually working in the large-scale limit,
to this traditionally sociological field. The sociologists themselves,
naturally, contribute to the development, and make the study of
communication networks a thriving interdisciplinary arena. Internet
has, of course, changed other aspects of society than science. In the
perennial human quest to find a partner, romantic or otherwise, the
Internet provides a new \textit{modus operandi}. In this paper we
study a Swedish Internet community, pussokram.com, intended for
romantic communication among adolescents. The methods we propose do
not only work for dating communities, but for Internet communities in
general. Our definition of an Internet community is a set of
HTML-pages facilitating some type of directed messages (i.e.\ not all
messages are broadcasted to the whole community), and where each
member is associated with a home page. This definition includes social
networking sites \cite{donathboyd} but excludes, for example, news
groups \cite{golder:ecomm}. The data we use, the same as in
Ref.~\cite{pok}, are anonymized lists of contacts---triples $(i,j,t)$
where $i$ is the ID-number of the sender, $j$ is the ID-number of the
receiver, and $t$ is the time (in seconds) since an arbitrary start
time. The IDs are integer numbers with no relation to the users of the
real community. We do not have access to the messages themselves,
nor the individual presentations on the user homepages (or any other
text or images either).

Researched data sets of electronic communication have
often been constructed from email exchange~\cite{adamic:email,%
bornholdt:email,eckmann:dialog,mejn:email,tyta:resp,tyler:email}, and
to a lesser extent from communication within Internet
communities~\cite{fiore:dating,pok,nioki}. The advantage with Internet
communities is that they are a closed system---all communication can
be recorded. The network of emails is a much larger and more
important phenomenon. But studies of email networks are plagued by
statistical biases due to the openness of the system. For email
communication one typically samples a set of individuals and
either restricts the data to messages sent within the group (which
leaves the message set incomplete)~\cite{eckmann:dialog} or one
includes contacts to outer vertices, which does not include contact
between outer vertices. For this reason the study of Internet
communities may not only be of interest \textit{per se}, but is also
illuminating towards the general structure of electronic communication.

The goal of this paper is to say something about what makes a user
successful in the community. Given our very restricted data---needless
to say, the most relevant information is in the text and imagery we do
not have---is there anything that can be said? We propose simple
measures for attractiveness and activity and observe that they are
positively correlated in the contact sequences. More than that, these
two quantities are more strongly correlated for the real-world data
than for the null model we propose. It thus pays off, directly or
(more likely) indirectly, to be an active community member.

\section{The community}

\begin{table}
  \caption{\label{tab:sizes}
    The sizes of our three data sets: The number of vertices $N$, edges
    $M$ and contacts $L$.
  }
  \begin{center}\begin{tabular}{l|ccc}
    & $N$ & $M$ & $L$ \\\hline
    guest book & 20,683 & 52,547 & 184,325\\
    messages & 21,537 & 50,938 & 264,819\\
    all contacts & 29,341 & 115,684 & 529,890\\
  \end{tabular}\end{center}
\end{table}

This study is based on data from the Swedish Internet community
pussokram.com, logged over 512 days (the same data as in
Ref.~\cite{pok}). Before the start of this data set
pussokram.com was a facility to send anonymous e-mails. Our data
starts at pussokram.com's beginning as a community. The number of active
users grew steadily during the sampling time. At the time
of writing, the community form of pussokram.com has ceased and the
original e-mail service has been re-established. The community form
(and its preceding and sequel e-mail service) was targeted at romantic
communication among youth; this was written quite explicitly in the
administrators' presentation and was conveyed throughout the HTML-pages
by text and iconography. This does not mean that all communication was
intended to lead to an offline encounter. In fact, a fraction of the
communication is probably regular chatting rather than
flirting~\cite{pok}.

There are four modes of contacts in the community:
\begin{enumerate}
\item Every user has a publicly accessible ``guest book'' where other
  users can post messages.
\item One can also send direct, e-mail-like, messages.
\item Each user has a list of ``friends'' at her (or his)
  homepage. For user $A$ to be listed as user $B$'s friend, first $A$
  has to send a ``friendship request'' to $B$ \ldots
\item \ldots and $B$ has to send a ``friendship acceptance'' to $A$.
\end{enumerate}

In this work we consider the sequences of guest book and message
contacts separately and the contact sequence of all four types of
contacts taken together. The number of vertices (persons) $N$, edges
(non-null dyads, i.e.\ pairs of vertices between which at least one
contact has occurred) $M$ and contacts (communications of one of the
four types) $L$ are listed in Table~\ref{tab:sizes}.

There are several ways for a member of pussokram.com to find others at
the community: 1. At a user homepage other similar users are
listed. 2. One can search others based on attributes like gender,
interests, place of residence, etc. 3. Pictures of $\sim 50$ users are
displayed on the login-page. 4. The ``friends'' of a user is listed at
user homepages. 5. The posters of guestbook messages are
displayed. 6. There is a lengthy interview with the ``user of the
month.'' A more detailed description of these means of finding others
is given in Ref.~\cite{pok}.

\section{Attractiveness and activity measures}

\begin{figure}
  \resizebox{\linewidth}{!}{\includegraphics{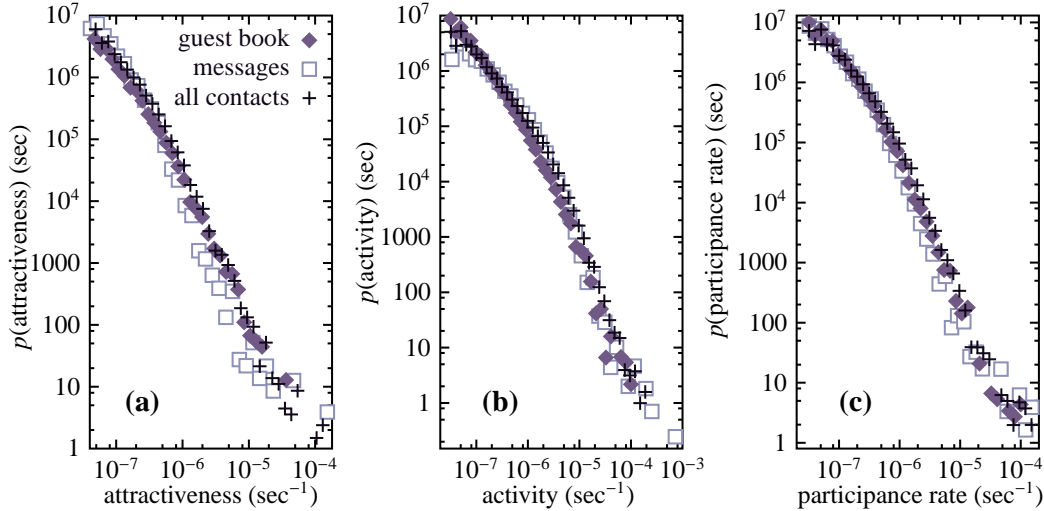}}
  \caption{\label{fig:hist} Logbinned probability density functions of
    attractiveness (a), activity (b) and participance rate.
  }
\end{figure}

Our measure of \textit{attractiveness} is the rate of incoming first
contacts, i.e.\ the number of contacts a member $i$ receives from
others (s)he have not been in contact with before divided by the total
time $t(i)$ that $i$ has been present in the data set. How one should
determine $t(i)$ in practice is not entirely trivial. One would like
$t(i)$ to be the time a vertex is a visible to the rest of the
community. The end-time is set by the end of the sampling time frame
$t_\mathrm{end}$. We chose the starting time for a vertex to be the
time (s)he sends or receives the first contact $t_1(i)$, so that
$t(i)=t_\mathrm{end}-t_1(i)$. Note that our $t(i)$ is an underestimate
of the real presence of a user in the data---one can always register
at the community (and thus become visible to others) before sending or
receiving any measure. We assume this time is negligible. More
elaborate estimates of $t_1$ would have to struggle with the fact a
member's behavior in the very beginning of his/her career hardly can
be estimated from the average behavior. The measure of
\textit{activity} we propose is the rate of any recorded contacts made
by a user, i.e.\ the number of contacts made by a member divided by
the same $t(i)$ as above. In addition to our activity and
attractiveness measures we also measure the \textit{participance
  rate}---the number of other members a member $i$ has contacted, or
been contacted by, divided by $t(i)$. This measures how fast an user
acquires connections to the rest of the community. Members with a high
participance rate will be the hubs of the social network generated by
the community interaction.

\begin{figure}
  \resizebox{\linewidth}{!}{\includegraphics{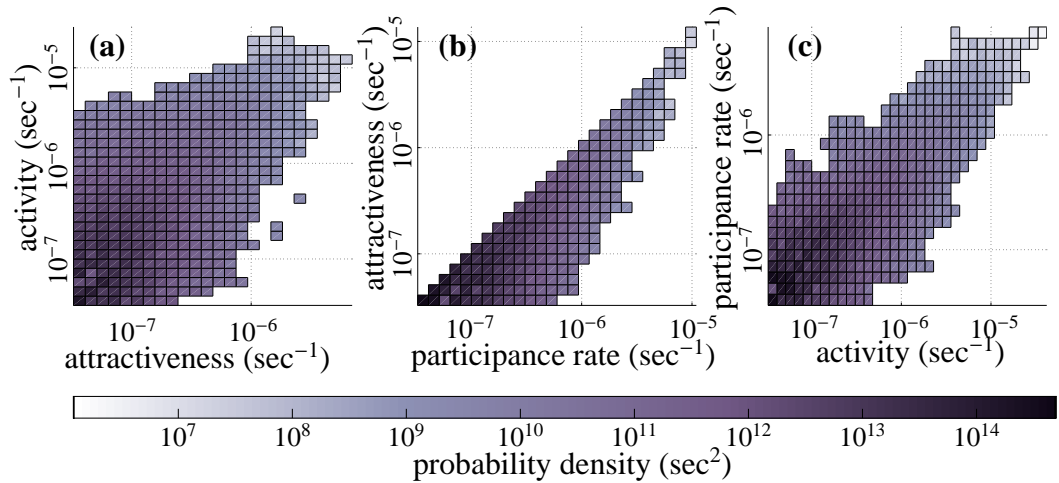}}
  \caption{\label{fig:dens} Logbinned probability density plots for the
    all-contacts data. (a) shows attractiveness vs.\ activity; (b)
    shows activity vs.\ participance ratio; and (c) shows participance
    ratio vs.\ attractiveness.
  }
\end{figure}

In Fig.~\ref{fig:hist} we plot the probability density functions for
our three quantities and our three data sets. We observe that all
three quantities are highly skewed. A broad distribution of activity
has been observed in Usenet communication~\cite{whit:usenet} and was
mentioned as a possible cause for the skewed degree distribution in
Ref.~\cite{pok}. Golder and Donath~\cite{golder:ecomm} discuss the
roles of participants in Usenet newsgroups; their classification is
partly based on the activity levels. For example, they call the most
active users ``celebrities,'' whereas ``newbies'' (new users) and
``lurkers'' (users who read but do not post messages) occupy the
low-end of the activity spectrum. The fact that the pussokram.com data
shows a very broad activity distribution is a sign that it is
meaningful to talk about ``celebrities'' and ``lurkers'' in this
community as well. The fat-tailed distributions of attractiveness and
participance rate are less trivial. We will discuss the origins of
these and their relation to the skewed activity distribution further
below. We note that the functional shape can be described as having
two different slopes in the log-log plots of Fig.~\ref{fig:hist}. This
is a rather common shape of probability density functions for social
and communication network data~\cite{my:rch,pok,mejn:scicolpre1}. We
will not dwell further on the details of the functional form.

Our three quantities are not independent. In Fig.~\ref{fig:dens} we
plot probability densities of the three pairs of measures. We see that
all three pairs of measures are positively correlated with each
other. I.e.\ an active user is probably also attractive; and has also,
likely, a high participance rate. By the definitions of attractiveness
and participance rate we note that the attractiveness is strictly
smaller than the participance rate, this is evident in
Fig.~\ref{fig:dens}(b). The colors of Fig.~\ref{fig:dens} represent
the average of the values of the four corners. Non-zero bins that do
not lie on a square with all corners being non-zero will not contribute
to these plots. For this reason the largest values of
Fig.~\ref{fig:hist} do not appear in Fig.~\ref{fig:dens}.

\section{Null model}

\begin{table}
  \caption{\label{tab:corr} The linear correlation coefficients
    between our quantities for the real-world data and the two null
    models. Numbers in parentheses are the standard errors in units of
    the last decimal place. Stars represent the correlation coefficient
    being significantly lower (* $p<0{.}05$; ** $p<0{.}01$) for the
    model networks than for the real-world network (except for the
    activity vs.\ participance rate correlations of model 2 where the
    model values are higher than for the real-world networks). The
    model network figures are averaged over 100 realizations.
  }
  \begin{center}\begin{tabular}{r|ccc}
    real world & attractiveness vs.\ activity & activity vs.\
    participance rate & part.\ rate vs.\  attr.\
    \\\hline
    guest book & 0{.}196 & 0{.}780 & 0{.}499 \\
    messages & 0{.}627 & 0{.}752 & 0{.}871 \\
    all contacts & 0{.}384 & 0{.}715 & 0{.}802 \\\hline\hline
   null model 1 & attractiveness vs.\ activity & activity vs.\
    participance rate & part.\ rate vs.\  attr.\
    \\\hline
    guest book &  0{.}059(4)* & 0{.}68(2) & 0{.}1104(7)** \\
    messages & 0{.}057(3)** & 0{.}71(2) &  0{.}103(3)**\\
    all contacts & 0{.}187(6)** & 0{.}68(2) & 0{.}279(5)** \\\hline\hline
   null model 2 & attractiveness vs.\ activity & activity vs.\
    participance rate & part.\ rate vs.\  attr.\
    \\\hline
    guest book & --0{.}0417(4)** & 0{.}9575(2)** &  0{.}0680(5)**\\
    messages & --0{.}0263(9)** & 0{.}9880(2)** &  0{.}025(2)**\\
    all contacts & --0{.}072(2)** & 0{.}9955(3)** &  0{.}016(3)**\\
  \end{tabular}\end{center}
\end{table}

In the previous section we defined three vertex-specific measures for
Internet community contact sequences. We measured the probability
distribution and the correlations between the measures. The design of
the measures does not exclude that correlations may be induced by the
growth of the community. To frame the correlations associated with the
psychology of the community members, we need something to
compare the values with, i.e.\ a null model. In fact, we propose two
null models, one more and one less restricted. We follow the general
approach of Ref.~\cite{katz:cug} and sample randomizations of the
real-world data set rather than constructing parametric null models.
In both these models we will keep the sizes $N$, $M$ and $L$ the same
as in the original contact sequences. Not only that, we assume
the growth of $N$ is unrelated to the psychology of the community
members, we also keep the time evolution of $N$ the same as in
the real data. Furthermore, as the community grows with time the
communication rate in the data should also be growing. With a
parametric null model this would be hard to implement, but we just
keep the set of times from the real data and the communication rate
exactly the same as in the real-world contact sequence.

In the first, less constrained, model we apply the restrictions
above. A randomized contact sequence is constructed as follows:
\begin{enumerate}
\item \label{enu:1c} For each vertex $i$ add one contact to or from
  (with equal probability) at time $t_1(i)$. The other vertex of
  the contact is chosen with uniform randomness among the vertices
  present in the data at this time.
\item \label{enu:rest} Draw $L-N$ times randomly
from the set of non-first contacts. (I.e.\ contacts $(i,j,t)$,
  such that $t\neq t_1(i)$ and $t\neq t_1(j)$.) Add
  contacts between vertices present in the community at these times.
\end{enumerate}
Step \ref{enu:1c} ensures that the number of active members is the
same in the randomized sequence as in the real-world data. Step
\ref{enu:rest} makes the set of times almost equal to the empirical set
of times. Note that, contacts between two vertices $i$ and $j$ at $t =
t_1(i) = t_1(j)$ will give rise to two contacts in the randomized
sequence; so the set of non-first contacts used in step \ref{enu:rest}
will be slightly ($<1\%$) larger than $L-N$.

As mentioned in the previous section a broad activity distribution has
been observed in many data sets similar to ours. One may argue that
activity is a rather independent trait, little connected to the
contact dynamics. Whether or not this is true, we construct a null
model to test how the reality differs from such a scenario. To
generate a random sequence realization for this model we loop over all
contacts $(i,j,t)$ and replace $j$ (the vertex the message is sent to)
by a random vertex $j'\neq i$ present in the community at time
$t$. Clearly this procedure conserves $L$. It may result in a few
isolated vertices (vertices of zero out-degree who looses their
incoming links), and a slightly altered time evolution of $N$
(vertices whose first contact is incoming may appear later in the
randomized community than in the real-world data). For our data these
effects are small, and we assume them to be negligible.

Values of the linear (Pearson's) correlation coefficient between
attractiveness, activity and participance rates for the real and null
model networks can be found in Table~\ref{tab:corr}. First we note
that the attractiveness and activity are more strongly correlated in
the real community than in both models, and that this holds for all
three networks. This means that the correlation between attractiveness
and activity is not an artifact of the growth of the community (by
comparison with model 1), neither is it a result of a skewed
distribution of a (hypothetical) intrinsic activity (by comparison
with model 2). Unlike the attractiveness vs.\ activity correlation,
the correlation between activity and participance rate seems to a
large extent to be an effect of the growth of the network. We see that
model 2 induces a very high correlation between activity and
participance rate. Model 2 randomizes the recipients but keeps the
rest of the communication the same as in the original data. The fact
that the users engage in dialogs is an explanation for the lower
correlation in the real world data---two users sending contacts to
only each other, but frequently is active but have low participance
ratio. The correlation between the participance rate and
attractiveness of the real data is even more different from the null
model than the attractiveness-activity correlation. This means that
the vertices of highest degree in the social network generated by the
communication are also the one that most frequently get incoming new
contacts. Another interesting observation is that the messages data
have higher attractiveness correlations than the guest book and all
contacts data. Since the messages are not publicly visible one can
assume that the real romantic communication takes place here rather
than in the guest book writing. This strengthen the conclusion that
active users are attractive in Internet dating. We note that (Spearman
type) rank correlations are typically a little stronger than the
linear correlations, but since they add little new information to the
discussion we do not include them in our tables.

\section{Summary and discussion}

In this paper we have introduced three measures to characterize users
in online communities: attractiveness, activity and participance rate
(measuring how fast a user get to know others in the community). While
these three measures are sensible in general Internet communities, we
evaluate them for an Internet dating community~\cite{pok}. How and why
people end up as partners is well studied in an off-line
setting~\cite{intimate,buss}; attractiveness in terms of body
shape~\cite{male:attr,female:attr}, facial
characteristics~\cite{face:attr} and body odor~\cite{odour:1} are all
well-known. With Internet communities playing an increasingly
important role on the dating scene~\cite{fiore:dating} one need  to
complement the studies of in-real-life attractiveness with
investigations of online romantic preferences. In this paper we  find
that the attractiveness (the number of incoming new contacts per time)
is positively correlated with activity (the number of contacts taken
per time), and that this correlation is significantly stronger than
for our two null models. We find this heightened correlation in the
real-world data even stronger in the dataset of e-mail-like messages
than in the other data sets (including messages visible to the whole
community). Since the activity, as we measure it, is rather invisible
to other members, this correlation has to be a secondary
effect. (Writing in guest books increases ones visibility and is a,
presumably small, direct effect.) It is natural to assume that the
users that send many messages, and thus invest much time into their
community, also spend comparatively more time embellishing their
homepages. The conclusion is thus that it pays off to spend effort on
one's personal presentation. Another factor is that highly active
users are most likely logged in more often than low-activity
users. Since currently logged in users are displayed, the increased
visibility of frequently logged in members may boost the
attractiveness of active users. We also find highly skewed,
power-law-like distributions of attractiveness, activity and
participance rate. The mechanisms behind electronic communication and
offline behavior are presumably very different. Nevertheless, similar
quantities in off-line relationships, such as the number of partners
per time, are also known have this feature~\cite{liljeros:sex}.

Research on the structure and dynamics of Internet communities is
still a young field. Their statistical advantages (being closed
systems) compared to e-mail exchange make us anticipate much future
work with both data analysis and modeling approaches.

\ack{The authors thank Mark Newman for illuminating
  discussions; Fredrik Liljeros for comments on the manuscript; and
  Christian Wollter and Michael Lokner for help with the data
  acquisition.}

\end{document}